\documentclass{article}
\usepackage{tikz}
\usetikzlibrary{shapes,arrows,arrows,positioning,fit}
\usepackage{multirow}
\usepackage{float}
\usepackage{spconf,amsmath,graphicx}
\usepackage[colorlinks=true,allcolors=red,linkcolor=red,anchorcolor=red,citecolor=red]{hyperref}

\title{A Fast and Accurate Pitch Estimation Algorithm Based on the Pseudo Wigner-Ville Distribution}
\name{  Yisi Liu$^{1}$, 
        Peter Wu$^{3}$,
        Alan W Black$^{2}$,
        Gopala K. Anumanchipalli$^{3}$}
\address{$^{1}$Department of Electronic and Information Engineering, 
        University of Chinese Academy of Sciences\\
       $^2$ Language Technologies Institute, Carnegie Mellon University \\ $^{3}$Berkeley Artificial Intelligence Research Lab, 
        University of California, Berkeley\\
        \small\texttt{liuyisi19@mails.ucas.ac.cn, peterw1@berkeley.edu, awb@cs.cmu.edu, gopala@berkeley.edu}}

\begin{document}
\maketitle
\begin{abstract}
Estimation of fundamental frequency (F0) in voiced segments of speech signals, also known as pitch tracking, plays a crucial role in pitch synchronous speech analysis, speech synthesis, and speech manipulation. In this paper, we capitalize on the high time and frequency resolution of the pseudo Wigner-Ville distribution (PWVD) and propose a new PWVD-based pitch estimation method. We devise an efficient algorithm to compute PWVD faster and use cepstrum-based pre-filtering to avoid cross-term interference. Evaluating our approach on a database with speech and electroglottograph (EGG) recordings yields a state-of-the-art mean absolute error (MAE) of around 4Hz. Our approach is also effective at voiced/unvoiced classification and handling sudden frequency changes.
\end{abstract}
\section{Introduction}
Fundamental frequency (F0), defined as the inverse of the vocal fold vibration period, is often estimated through the pitch tracking task. Reliable pitch tracking has a wide range of applications, including speech synthesis \cite{synthesis_1}, speech prosody analysis \cite{analysis_1}, speech manipulation \cite{manipulation_1, STRAIGHT}, melody extraction \cite{melody}, glottal source processing \cite{glottal}, and intonation teaching \cite{intonation_teaching}.

Traditional methods for pitch estimation can generally be split into three categories. First is the cepstrum-based method \cite{cepstrum_1, cepstrum_2}, where large peaks in the cepstrum correspond to pitch periods in the frequency domain. Second is the correlation-based method, which can be viewed in terms of three sub-categories: (a) auto-correlation-based \cite{autocorrelation}, where the auto-correlation of the output of a spectrum flattener is utilized; (b) normalized-cross-correlation-based \cite{RAPT}, which is also the key algorithm for REAPER; and (c) YIN-based \cite{STRAIGHT, YIN, pYIN}, where the cumulative mean normalized difference function is used in addition to auto-correlation function. The third category of traditional methods leverage time-frequency representations, among which wavelets \cite{wavelet} are the most popular. 

In recent years, researchers have proposed data-driven algorithms for pitch estimation. Previously, machine learning methods were unable to outperform traditional approaches \cite{pYIN} due to a lack of annotated data \cite{SPICE}. CREPE \cite{CREPE} circumvented this constraint and achieved state-of-the-art results by training on a synthetically generated dataset for F0 tracking \cite{syn_dataset}. Later, SPICE \cite{SPICE} used self-supervised learning to obtain pitch estimation results comparable to CREPE.

So far, among traditional signal processing methods, those based on time-frequency representations have not gained much attention. One primary reason for this is that time and frequency resolutions are restricted by the Heisenberg uncertainty principle \cite{Mallat}. Specifically, one cannot increase time and frequency resolutions simultaneously without introducing artifacts. For traditional short-time Fourier transform (STFT) and wavelet transforms, more time support means worse time resolution and finer frequency resolution, and vice versa.

The Wigner-Ville distribution (WVD) has high resolution in both the time and the frequency domains, making it one of the most powerful and fundamental time-frequency representations~\cite{Mallat, WVD_1, WVD_2}. Despite these remarkable properties, WVD has not been widely used, since: (1) WVD is computationally expensive; (2) the existence of interference (cross terms) among different components may introduce false information; and (3) WVD is highly non-local.

In this paper, we overcome the drawbacks of WVD and present a high-performance pitch tracker based on the pseudo WVD (PWVD). Inspired by \cite{efficient_WVD}, we utilize the Hilbert transform, downsampling, and segmentation as well as implement the WVD in terms of the fast Fourier transform (FFT) in order to compute the PWVD faster. Additionally, we use cepstrum-based pre-filtering to eliminate cross terms between F0 and its multiples. Finally, PWVD is used to make our pitch tracker more sensitive to frequency changes. We show that our algorithm outperforms the widely used REAPER and STRAIGHT methods, and is comparable to or better than the state-of-the-art approaches like pYIN and CREPE, in terms of both mean absolute error (MAE) and F0 Frame Error (FFE).
\section{Method}
\label{sec:method}
This section describes our proposed PWVD-based pitch tracker step by step to show what makes it efficient and effective. First, we present the basic theory of WVD. Then, we introduce a five-stage pipeline, which includes downsampling, voiced/unvoiced (V/UV) classification, segmentation, cepstrum-based pre-filtering, and PWVD, in order to show the implementation of our method.
\subsection{Basics of WVD}
\begin{figure}[tb]
        \begin{minipage}[b]{1.0\linewidth}
        \centering
        \centerline{\includegraphics[width=9.8cm]{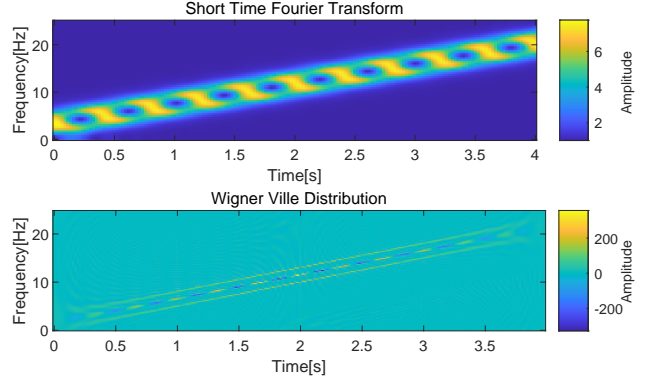}}
        \end{minipage}
        \caption{The STFT and WVD of \eqref{sum of chirps}, where sampling frequency $f_s$ = 50Hz, $k$ = 5Hz/s, $f_0$ = 3Hz, and $s(t)$ resides in 0 to 4s. A Hann window of 32 points is used for STFT, with a stride (overlap) of 31 points and an FFT length of 128 points. STFT suffers greatly from spectrum smearing and artifacts as a result of low resolution. For WVD, the time-frequency information is perfectly captured by two parallel thin straight lines with high resolution, but there are also oscillating parts, i.e., cross terms, due to interference between the two lines.}
        \label{fig:stft_wvd}
    \end{figure}

    The Wigner-Ville distribution of signal $x(t)$ is given by
    \begin{equation}
        W_x(t,\omega) = \int_{-\infty}^{+\infty}x(t+\frac{\tau}{2})x^*(t-\frac{\tau}{2})e^{-j\omega\tau}d\tau.\label{WVD}
    \end{equation}
    Defining the instantaneous auto-correlation function $R(t,\tau)$ as
    \begin{equation}
        R(t,\tau) = x(t+\frac{\tau}{2})x^*(t-\frac{\tau}{2}),
    \end{equation}
    we note that $W_x(t,\omega)$ can be viewed as the Fourier Transform of $R(t,\tau)$ along the $\tau$ axis, i.e.,
    \begin{equation}
        W_x(t,\omega) = \int_{-\infty}^{+\infty}R(t,\tau)e^{-j\omega\tau}d\tau.
    \end{equation}
    We utilize this property in our FFT step in Section \ref{sec:stage_5_pwvd}. One standard way of defining the discrete version of WVD is
    \begin{equation}
        W_x[n,k] = \sum_{m = -N+1}^{m = N-1}x[n+\frac{m}{2}]x^*[n-\frac{m}{2}]e^{-j\frac{2\pi}{N}km}.
    \end{equation}
    This definition involves half-integer indices, which are computed in the MATLAB built-in function \texttt{wvd} using interpolations. In Section \ref{sec:stage_5_pwvd}, we introduce a more efficient algorithm for computing $W_x[n,k]$ that avoids interpolations. When there are multiple components in a signal $s(t)$, e.g., 
    \begin{equation}
        s(t) = s_1(t) + s_2(t)\label{s},
    \end{equation}
    the WVD of $s(t)$, i.e., $W_s(t,\omega)$ in \eqref{WVD}, will contain cross terms $I(s_1,s_2)$ given by
    \begin{equation}
    \begin{split}
        I(s_1,s_2) &= \int_{-\infty}^{+\infty}s_1(t+\frac{\tau}{2})s_2^*(t-\frac{\tau}{2})e^{-j\omega\tau}d\tau\\
                   &+ \int_{-\infty}^{+\infty}s_2(t+\frac{\tau}{2})s_1^*(t-\frac{\tau}{2})e^{-j\omega\tau}d\tau.
    \end{split}
    \end{equation}
    The existence of cross terms does not mean that the original signal has energy distributed in the cross term's time-frequency neighborhood. In this sense, $I(s_1,s_2)$ can be considered as artifacts that need to be removed \cite{Cohen}. An example of cross terms and WVD's high time-frequency resolution can be found in Figure \ref{fig:stft_wvd}, where we compare the WVD and STFT of the following sum of two chirp signals:
    \begin{equation}\label{sum of chirps}
        s(t) = \cos(\pi kt^2 + 2\pi f_0t) + \cos(\pi kt^2).
    \end{equation}

\subsection{Stage 1: Downsampling}
    Normally, for human speakers, pitch approximately stays between 60Hz and 400Hz. Based on this assumption, we filter out frequency components outside of this range. We then downsample the band-passed signal to decrease the sampling frequency to $f_d$ = 800Hz and refer to the resulting signal as $x_d$. This downsampling operation noticeably reduces the amount of computation for PWVD.
\subsection{Stage 2: V/UV Classification}
    Voiced sounds are quasi-periodic, while unvoiced sounds are more similar to noise and contain much higher frequency components. Based on the approach in~\cite{V/UV}, we use energy thresholding for voiced/unvoiced (V/UV) classification. Essentially, the voiced areas of $x_d$ generally have much higher energy than the unvoiced areas. We check the average energy of $x_d$ frame by frame and split higher-energy frames into subframes for more precise results. Here, we choose a frame size of 25ms, and a threshold of 0.2E, where E is the average energy of $x_d$.
\subsection{Stage 3: Segmentation}     
    In this stage, we further divide each of the voiced areas into segments of around 150ms, i.e., 120 points at $f_d$ = 800Hz. We base this segmentation operation on the assumption that shorter periods of time will make drastic frequency changes less likely to occur. This helps to reduce the appearance of cross terms and prepares the signal for the subsequent pre-filtering stage.
\subsection{Stage 4: Cepstrum-based Pre-filtering}\label{pre-filtering}
    The purpose of pre-filtering is to filter out harmonic structures so that approximately only the fundamental frequency component remains, reducing most cross terms for the upcoming PWVD. In order to do this, the average F0 ($f_a$) of each voiced segments is required. Here, we combine cepstrum with spectrum to extract $f_a$. First, we pass the pre-downsampling voiced segments (here sampled at $f_s$ = 16kHz) into a low-pass filter (LPF) with stopband edge frequency 1kHz. Second, we do cepstrum maximum detection on the filtered segments to calculate $f_{raw}$: 
    \begin{equation}
        f_{raw} = \frac{f_s}{\tau},
    \end{equation}
    where $\tau$ is the quefrency index where the cepstrum reaches its maximum within the frequency range of 80Hz to 320Hz. Third, we generate $f_a$ candidates as multiples and factors of $f_{raw}$, also within the frequency range of 80Hz to 320Hz. Finally, we output the smallest candidate that has a prominent spectrum peak in its frequency neighborhood.
    
    After extracting the average F0 values, we concatenate the respective neighboring points to each of the voiced segments in order to mitigate the tendency of WVD to fade at the edges of a signal. Figure \ref{fig:stft_wvd} contains an example of such unwanted edge behavior. Following this concatenation step, we apply pre-filtering on each voiced segment to just keep 0.7$f_a$ to 1.4$f_a$ frequency components.
\subsection{Stage 5: PWVD}
\label{sec:stage_5_pwvd}
    \begin{table*}[tb]
        \centering
        \begin{tabular}{p{1.9cm}||c|c|c||c|c|c||c|c|c}
        \hline\hline
        Datasets & \multicolumn{3}{c||}{bdl} & \multicolumn{3}{c||}{jmk} & \multicolumn{3}{c}{slt}\\
        \hline
        Metrics & MAE & PVE & FFE & MAE & PVE & FFE & MAE & PVE & FFE\\
        \hline\hline
        PWVD & \textbf{3.12} & 5.79 & 9.80 & \textbf{3.88} & 8.61 & \textbf{9.83} & 6.36 & 11.13 & 7.82\\
        \hline
        REAPER & 3.35 & 6.17 & 12.42 & 4.39 & 9.06 & 12.52 & 6.33 & 10.73 & 10.98\\
        \hline
        STRAIGHT & 4.88 & 24.43 & \textbf{8.19} & 7.21 & 28.19 & 15.06 & 7.48 & 23.45 & 9.14\\
        \hline
        pYIN & 3.21 & \textbf{4.93} & 16.15 & 4.29 & \textbf{7.44} & 10.25 & \textbf{6.10} & 10.08 & 9.15\\
        \hline
        CREPE & 3.30 & 4.99 & 10.06 & 3.99 & 7.71 & 9.84 & 6.30 &\textbf{ 9.97} & \textbf{7.69}\\
        \hline\hline
        \end{tabular}
        \caption{Mean absolute error (MAE), Pitch Variation Error (PVE) and F0 Frame Error (FFE) across 3 datasets.}
        \label{tab:pitch}
    \end{table*}
    
    \begin{figure}[tb]
        \begin{minipage}[b]{1.0\linewidth}
        \centering
        \centerline{\includegraphics[width=9.8cm]{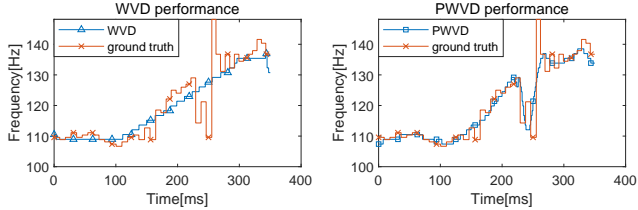}}
        \end{minipage}
        \caption{A comparison between WVD and PWVD pitch tracking results of ``Your letter". A hann window of 40ms is chosen for PWVD.}
        \label{fig:wvd_pwvd}
    \end{figure}
    The pre-filtered voiced segments are passed into this stage. Recall \eqref{WVD}, where interpolation is involved to compute WVD. Similar to the idea in \cite{efficient_WVD}, where the Hilbert transform is utilized to avoid spectrum aliasing, we can modify \eqref{WVD} into
    \begin{equation}\label{better WVD}
        \begin{split}
            W_x[n,k] &= \sum_{m = -N+1}^{m = N-1}\Tilde{x}[n+m]\Tilde{x}^*[n-m]e^{-j\frac{4\pi}{N}km}\\
                     &= \sum_{m = -N+1}^{m = N-1}\hat{R}[n,m]e^{-j\frac{4\pi}{N}km}\\
                     &\sim FFT(\hat{R}) + FFT^*(\hat{R}) - |\Tilde{x}[n]|^2.
        \end{split}
    \end{equation}
    Here $\Tilde{x}$ is the analytic signal of $x$ and $*$ means complex conjugate. $4\pi$ instead of $2\pi$ in the exponent just represents frequency stretching, i.e., the 3rd line is calculated first, and then all frequencies are divided by 2. We note that interpolation is avoided, and WVD is converted into an FFT computation.

    After computing WVD, we estimate F0 by looking for the first prominent peak of WVD at each time index $n$. Figure \ref{fig:wvd_pwvd} shows an example of pitch tracking using WVD, which, although functional, fails to track the ground truth closely enough. This is due to the non-local nature of the WVD calculation, which assigns equal weights to the past, present, and future points. If we apply a window function $w[m]$ to $\hat{R}[n,m]$ to emphasize the ``present", this non-local effect can be alleviated. This operation is called the Pseudo-WVD~\cite{Cohen}:
    \begin{equation}
         PW_x[n,k] = \sum_{m = -N+1}^{m = N-1}\hat{R}[n,m]w[m]e^{-j\frac{4\pi}{N}km}.
    \end{equation}
    Figure \ref{fig:wvd_pwvd} compares pitch tracking using PWVD with the WVD method. Compared to WVD, PWVD is more sensitive to frequency changes as a result of emphasizing the ``present".
\section{Results}
\subsection{Datasets}
    \begin{figure}[tb]
        \begin{minipage}[b]{1.0\linewidth}
        \centering
        \centerline{\includegraphics[width=9.8cm]{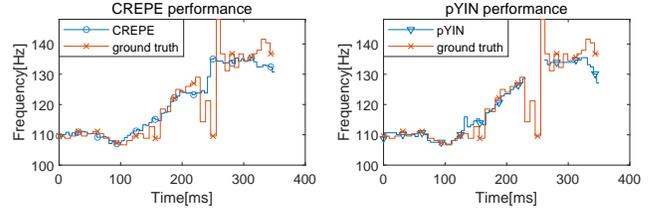}}
        \end{minipage}
        \caption{CREPE and pYIN pitch tracking on the same utterance ``Your letter" appearing in Figure \ref{fig:wvd_pwvd}. The gap in pYIN's prediction is due to pYIN classifying that region as unvoiced.}
        \label{fig:crepe_pyin}
    \end{figure}
    We conduct our experiments with the CMU ARCTIC databases \cite{database}. Each single-speaker speech database contains nearly 1150 phonetically balanced English sentences. We choose the speakers bdl (US male), jmk (Canadian male) and slt (US female) for evaluating pitch tracking, since they have EGG signals recorded simultaneously with clean speech signals under studio conditions. The F0 and V/UV ground truths are extracted from differentiated EGG signals.
\subsection{Error Metrics}
    Three error metrics were chosen for comparison: mean absolute error (MAE), pitch variation error (PVE) and F0 frame error (FFE) \cite{metrics}. MAE describes the average absolute error between the F0 ground truth and the estimated F0. We calculate MAE within the intersection of voiced areas given by the ground truth and different pitch estimation algorithms. PVE is the standard deviation of the absolute error, which describes how steadily a pitch tracker performs. FFE is the proportion of frames for which an error is observed. Here, the error includes Gross Pitch Error (GPE, the estimated pitch differs from the ground truth by more than 20\%) and Voicing Decision Error (VDE, voiced frames classified as unvoiced or unvoiced frames classified as voiced). FFE can be seen as a single metric to evaluate the overall performance of a pitch tracker.
    
\subsection{Time Efficiency}
    \begin{figure}[tb]
        \begin{minipage}[b]{1.0\linewidth}
        \centering
        \centerline{\includegraphics[width=10cm]{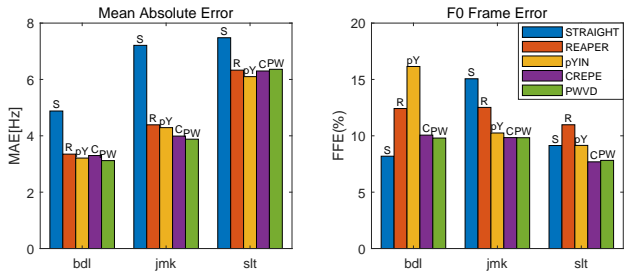}}
        \end{minipage}
        \caption{Bar charts for mean absolute error (MAE) and F0 Frame Error (FFE). The initials of methods are placed at the end of each bar.}
        \label{fig:bar}
    \end{figure}
    The built-in \texttt{wvd} of MATLAB takes around 60s to process 0.5s of a signal sampled at $f_s$ = 16kHz. Moreover, once the duration of the input signal exceeds 1s at $f_s$ = 16kHz, memory shortage will stop the execution. In comparison, with the proposed algorithm mentioned in Section \ref{sec:method}, a runtime evaluation over the three databases (over 3000 utterances) yielded a result of 0.665RT. I.e., it takes about 0.665s to process 1s of a signal on average, which is much more time efficient.
\subsection{Evaluation}
    We compare our proposed PWVD method with 4 popular pitch trackers: REAPER \cite{RAPT}, STRAIGHT \cite{STRAIGHT}, pYIN \cite{pYIN} and CREPE \cite{CREPE}. Table \ref{tab:pitch} and Figure \ref{fig:bar} summarize the results. Since CREPE does not support V/UV classification, our experiments with CREPE use our V/UV method. For MAE, our proposed PWVD-based pitch tracker performs the best in the datasets bdl and jmk, with the lowest MAE and a p value less than $10^{-10}$, indicating a fairly accurate tracking performance. For the overall performance metric FFE, PWVD has the lowest FFE of 9.83\% in the dataset jmk, which is significantly lower than REAPER, STRAIGHT and pYIN with a p value less than $10^{-10}$, and lower than CREPE with a p value of 0.038, obtaining a state-of-the-art overall performance. With the datasets bdl and slt, PWVD has the second lowest FFE, which is also noticeably lower than the other three pitch trackers. PWVD also has medium-level PVE, suggesting a steady performance with relatively few halving or doubling errors due to pre-filtering. Figure \ref{fig:crepe_pyin} contains an example of CREPE's and pYIN's performance. Compared to PWVD (Figure \ref{fig:wvd_pwvd}), CREPE and pYIN fail to track sudden frequency changes closely enough. Moreover, pYIN tends to classify more voiced areas as unvoiced, which causes its high FFE.
\section{Discussions and Conclusion}
    We present a pitch tracker based on the pseudo Wigner-Ville distribution (PWVD) that utilizes the high time-frequency resolution of PWVD. Additionally, we devise an algorithm that calculates WVD much faster than the previous implementation and use cepstrum-based pre-filtering to eliminate most cross terms. We also utilize PWVD in order to make WVD more sensitive to sudden frequency changes. An evaluation of five pitch trackers on three selected datasets yielded low MAE and FFE for the proposed pitch tracker, obtaining state-of-the-art performance with multiple dataset-metric combinations.

    Since PWVD is calculated in the voiced areas only, a more precise V/UV classifier can potentially improve the performance of $f_a$ extraction system mentioned in \ref{pre-filtering}, and therefore reduce halving error and doubling error. Moreover, considering the estimated F0 of adjacent points when proposing F0 candidates for present point may enforce temporal smoothness, reducing abrupt frequency changes in the output.
\section{Acknowledgements}
    This research is supported by the following grants to PI Anumanchipalli --- NSF award 2106928, Google Research Scholar Award, Rose Hills Foundation and Noyce Foundation.  
\bibliographystyle{IEEEbib}
\bibliography{refs}
\end{document}